\documentclass[aps,prc,groupedaddress,manuscript]{revtex4}

\usepackage{graphicx}

\begin{document}

\title{A model comparison of resonance lifetime modifications, a soft equation of state and non-Gaussian effects on $\pi-\pi$ correlations at FAIR/AGS energies}

\author {Qingfeng Li$\, ^{1,2}$\footnote{E-mail address: liqf@fias.uni-frankfurt.de} and Marcus Bleicher,$\, ^{3}$}
\address{
1) Frankfurt Institute for Advanced Studies (FIAS), Johann Wolfgang
Goethe-Universit\"{a}t, Max-von-Laue-Str.\ 1, D-60438 Frankfurt am
Main, Germany \\
2) School of Science, Huzhou Teachers College, Huzhou 313000,
China\\
3) Institut f\"{u}r Theoretische Physik, Johann Wolfgang
Goethe-Universit\"{a}t, Max-von-Laue-Str.\ 1, D-60438 Frankfurt am
Main, Germany\\
4) Gesellschaft f\"ur Schwerionenforschung (GSI), Planckstr. 1,
D-64291 Darmstadt, Germany
 }


\begin{abstract}
HBT correlations of $\pi^--\pi^-$ pairs at FAIR/AGS energies are
investigated by using the UrQMD transport model and the CRAB
analyzing program. Three different possible sources (treatment of
resonance lifetimes, a soft equation of state and non-Gaussian
effects) to understand the HBT $R_O/R_S$ puzzle are investigated.
Firstly, we find that different treatments of the resonance decay
time can not resolve the HBT time-related puzzle, however it can
modify the HBT radii at low transverse momenta to some extent to
explain the data slightly. Secondly, with a soft equation of state
with momentum dependence, the measured transverse momentum dependent
HBT radii and $R_O/R_S$ ratio can be described fairly well. Thirdly,
non-Gaussian effects are visible in the calculated correlation
function. Using the Edgeworth expansion, one finds that the
non-Gaussian effect is strongest in the longitudinal direction and
weakest in the sideward direction.

(Some figures in this article are in colour only in the electronic
version)
\end{abstract}



\maketitle

\section{Introduction}
It is well-known that one can extract information on the space-time
dimensions of the particle emission source (the region of
homogeneity) in heavy-ion collisions (HICs) by using the
Hanbury-Brown-Twiss interferometry (HBT)
\cite{HBT54,Goldhaber60,Bauer:1993wq} techniques. With the ongoing
advances in detectors and accelerators one is able to explore
collision energies for HICs from less than $\sqrt {s_{NN}} \sim
2.5$~GeV (SIS/FAIR energy regime), $2.5- 20$~GeV (FAIR/AGS and SPS)
up to $20 - 200$ GeV (RHIC). Within this broad energy region Quantum
Chromodynamics (QCD) predicts a transition from a hadron gas to a
quark-gluon plasma (QGP), and it is expected that this new QGP state
of matter exists at least temporarily in the center of HICs. During
the phase transition (i.e., in the mixed phase), it was proposed
that a nontrivial transition in the spatio-temporal characteristics
of the source exists \cite{Rischke:1996em}. Unfortunately, so far
the excitation functions of the HBT parameters have not shown any
{\it obvious} discontinuities in experiments with energies from SIS,
AGS, SPS, up to RHIC \cite{Lisa:2005dd}. Instead, several unexpected
and interesting phenomena occurred, namely, the ``E-puzzle'', the
``t-puzzle'', and the ``non-Gaussian'' effect
\cite{Soff:2000eh,Lisa:2005dd,Li:2007im,Wiedemann:1999qn} (for an explanation of
these terms, see below). Objectively speaking, the existing
theoretical investigations are still in-sufficient and further
thorough investigations are needed.

In a recent work on the HBT correlation, adopting the
Ultra-relativistic Quantum Molecular Dynamics (UrQMD, v2.2)
transport model \cite{Bass98, Bleicher99,Bratkovskaya:2004kv} and
the ``correlation after-burner'' (CRAB, v3.0$\beta$) analyzing
program \cite{Koonin:1977fh,Pratt:1994uf,Pratthome}, the transverse
momentum, system-size, centrality, and rapidity dependence of the
HBT parameters of the sources of different identical particle pairs
(two $\pi$s, two $K$s, and two $\Lambda$s) at AGS, SPS and RHIC
energies \cite{lqf2006,lqf20062,Li:2007im,Li:2007yd,Li:2008ge} were
investigated. It was found that although the calculations are
generally in line with the experimental data, discrepancies are not
negligible. One of the most puzzling phenomena is that the
calculated ratio of HBT radii in the outward direction ($R_O$) and
in sideward direction ($R_S$) from central HICs is always larger
than that extracted from the data at all investigated energies, if
the cascade mode is employed, which was named as the HBT
time-related puzzle (``HBT t-puzzle'') \cite{Li:2007im}. After
considering a soft equation of state (EoS) with momentum dependence
(dubbed as SM-EoS) for formed baryons and a simple Skyrme-like
(density dependent) potential for the ``pre-formed'' particles
\cite{Li:2007yd}, the HBT radius $R_O$ is pushed down and the $R_S$
is pulled up to approach the data so that the ``HBT t-puzzle''
disappears throughout the whole energy region. Meanwhile, the
transverse mass ($m_T$) scaling, which was predicted in Ref.\
\cite{Csorgo:1995bi}, and has been probed by several experiments
recently \cite{Lisa:2005dd,Bearden:2001sy}, can also be much better
understood with the help of ``pre-formed'' particle potentials in
HICs \cite{Li:2008ge}. Therefore, it was concluded that the
interaction (here it is embodied with potentials) of particles at
the early stages of HICs can help to solve the ``HBT t-puzzle''.

However, besides the potential effects, the resonance dynamics and
decay may also influence the momentum distribution of emitted
particles as well as their correlation \cite{Wiedemann:1999qn}.
Since this question has been left aside in our previous works, in
this paper, we want to complete the discussion by investigating the
effects of the handling of resonance decay times on the HBT
quantities by adopting the microscopically transport model UrQMD.
The different contributions from the resonance decay and from the
potential modification are then compared. In order to have a clearer
situation, in this paper we restrict ourselves to the low energy
region, i.e., the FAIR/AGS energy regime. In this energy region
hadronic interactions dominate the dynamics. String dynamics is
negligible and the quarks are still confined and the interactions
between them do not need to be taken into account during the HICs. A
further simplification is that the available resonances at these
beam energies are mainly the $\Delta(1232)$s, and one can restrict
the analysis of resonance modification to this hadron. Due to the
fact that pions have the largest abundance, they are well suited for
the present analysis. Negatively charged pions can be easily
measured (no contamination from misidentified K$^+$s and protons)
and experimental data are therefore available from most experiments
today. Thus, we focus the present analysis on negatively charged
pions.

The paper is arranged as follows. In the next section, the UrQMD
transport model and the CRAB analyzing program are introduced. The
different treatments of $\Delta$ decay and the effects on pion
production are also discussed. In Section 3, firstly, the Gaussian
fitting to the one- and the three-dimensional correlation functions
of negatively charged pions and the non-Gaussian effect are explored
and discussed. Secondly, we show the comparison of the transverse
momentum dependence of HBT radii and $R_O/R_S$ ratios (from the
Gaussian fitting) between calculations with different treatments of
the resonance decay and with(without) mean field potentials and
experimental data at AGS energies. Finally, in section 4, a summary
and outlook is given.

\section{models and treatments}
\subsection{UrQMD transport model}
The first formal version (ver1.0) of the UrQMD transport model was
published in the end of last century  \cite{Bass:1997xw,Bass98,
Bleicher99}. Since then, a large number of successful theoretical
analyses, predictions, and comparisons with data based on this
transport model have been accomplished for {\it pp}, {\it p}A and AA
reactions for a large range of beam energies, i.e., from SIS, AGS,
SPS, up to RHIC \cite{urqmdweb1}.

The UrQMD model is based on analogous principles as the Quantum
Molecular Dynamics model (QMD)
\cite{Aichelin:1986wa,Aichelin:1991xy} and the Relativistic Quantum
Molecular Dynamics model (RQMD) \cite{Sorge:1989dy}. Similar to QMD,
hadrons are represented by Gaussian wave packets in phase space, and
the phase space of hadron $i$ is propagated according to Hamilton's
equation of motion:

\begin{equation}
{\bf \dot{r}}_i=\frac{\partial H}{\partial {\bf p}_i},
\hspace{1.5cm} \mathrm{and} \hspace{1.5cm} {\bf
\dot{p}}_i=-\frac{\partial H}{\partial {\bf r}_i}. \label{Hemrp}
\end{equation}
Here ${\bf {r}}$ and ${\bf {p}}$ are the coordinate and momentum of
hadron $i$. The Hamiltonian $H$ consists of the kinetic energy $T$
and the effective two-body interaction potential energy $V$,

\begin{equation}
H=T+V, \label{Hamifunc}
\end{equation}
and
\begin{equation}
T=\sum_i (E_i-m_i)=\sum_i (\sqrt{m_i^2+{\bf p}_i^2}-m_i). \label{HT}
\end{equation}
In the standard version of UrQMD model \cite{Bass:1997xw,Bass98,
Bleicher99}, the potential energies include the two-body and
three-body (which can be approximately written in the form of
two-body interaction) Skyrme- (also called as the density dependent
terms), Yukawa-, Coulomb-, and Pauli-terms as a base.

\begin{equation}
V=V_{sky}^{(2)}+V_{sky}^{(3)}+V_{Yuk}+V_{Cou}+V_{Pau}. \label{HV}
\end{equation}
And the single particle potential follows from $U= \delta V/\delta
f$, where $f$ is the phase space distribution function which reads
as

\begin{equation}
f({\bf r},{\bf p})=\sum_i f_i({\bf r},{\bf p})=\sum_i \frac{1}{(\pi
\hbar)^3}e^{-({\bf r}-{\bf r}_i)^2/2L^2}e^{-({\bf p}-{\bf
p}_i)^2\cdot 2L^2/\hbar^2}, \label{wigfunc}
\end{equation}
here $L$ is the width parameter of the wave packet. Recently, in
order to be more successfully applied into the intermediate energy
region ($0.1\lesssim E_b \lesssim 2$A GeV), the UrQMD has contained
more potential terms \cite{Li:2005gf}, those are, 1) the
density-dependent symmetry potential, which is essential for
isospin-asymmetric reactions at intermediate and low energies, and
2) the momentum-dependent term \cite{Bass:1995pj}
\begin{equation}
U_{md}=t_{md} \ln^2[1+a_{md}({\bf p}_i-{\bf p}_j)^2]\rho_i/\rho_0,
\label{umd}
\end{equation}
where $t_{md}$ and $a_{md}$ are parameters, $\rho_0$ is the normal
density. $\rho_i$ is the density of the baryon $i$,

\begin{equation}
\rho_i=\int\rho(\bf{r_i}) \rho d\bf{r}=\int\rho(\bf{r}_i)
\sum_j\rho(\bf{r}_j) d\bf{r}=\frac{1}{(4\pi
L)^{3/2}}\sum_{j}e^{-\frac{(\bf{r}_i-\bf{r}_j)^2}{4 L}}.
\label{rhoi}
\end{equation}
With these updates, some sensitive probes of the (density dependent)
symmetry potential have been proposed
\cite{Li:2005zz,Li:2005gf,Li:2005kq,Li:2006wc}. Further, it was also
found that the experimental pion and proton directed and elliptic
flows from HICs with beam energies from $\sim 100$A MeV to $~2$A GeV
can be well described \cite{Petersen:2006vm}.

At higher beam energies, the Yukawa-, Pauli-, and symmetry-
potentials of baryons becomes negligible, while the Skyrme- and the
momentum-dependent part of potentials still influence the whole
dynamical process of HICs. With the help of a covariant prescription
of mean field from RQMD/S \cite{Maruyama:1996rn}, the effects of the
mean field with momentum dependence on collective flows from HICs at
2-158A GeV energies were studied by a Jet AA Microscopic
Transportation Model (JAM) and it has been found that the momentum
dependence in the nuclear mean field is important for the
understanding of the proton collective flows at AGS and even at SPS
energies \cite{Isse:2005nk}. In this work, we choose the same SM-EoS
as that in Ref.~\cite{Isse:2005nk}, where the momentum dependent
term reads as

\begin{equation}
U_{md}=\sum_{k=1,2}\frac{t_{md}^k}{\rho_0}\int d{\bf
p}_j\frac{f({\bf r}_i,{\bf p}_j)}{1+[({\bf p}_i-{\bf
p}_j)/a_{md}^k]^2}, \label{vmdnew}
\end{equation}
where the $U_{md}$ consists of two parts with separate parameters in
order to fit better the real part of the optical potential
\cite{Hama:1990vr}. We found that the momentum dependent term is
also important to explain the transverse momentum dependence of the
HBT parameters at AGS energies \cite{Lisa:2000hw,Li:2007yd}.
Furthermore, as in Ref.\ \cite{Isse:2005nk}, the relativistic
effects on the relative distance ${\bf r}_{ij}={\bf r}_i-{\bf r}_j$
and the relative momentum ${\bf p}_{ij}={\bf p}_i-{\bf p}_j$
employed in the two-body potentials (Lorentz transformation) are
considered as follows:

\begin{equation}
{\bf \tilde{r}_{ij}^2}={\bf r}_{ij}^2+\gamma_{ij}^2({\bf
r}_{ij}\cdot {\bf \beta}_{ij})^2, \label{rboost}
\end{equation}
\begin{equation}
{\bf \tilde{p}_{ij}^2}={\bf
p}_{ij}^2-(E_i-E_j)^2+\gamma_{ij}^2(\frac{m_i^2-m_j^2}{E_i+E_j})^2.
\label{pboost}
\end{equation}
In Eqs.~\ref{rboost} and \ref{pboost} the velocity factor ${\bf
\beta}_{ij}$ and the corresponding $\gamma$-factor of $i$ and $j$
particles are defined as $ {\bf \beta}_{ij}=({\bf p}_i+{\bf
p}_j)/(E_i+E_j)$ and $ \gamma_{ij}=1/\sqrt{1-{\bf \beta}_{ij}^2}$.

A covariance-related reduction factor for potentials in the
Hamiltonian, $m_j/E_j$, was introduced in the simplified version of
RQMD model \cite{Maruyama:1996rn} and adopted in this work as well.

The collision term of the UrQMD model treats 55 different baryon
species (including nucleon, $\Delta$, $\Lambda$, $\Sigma$, $\Xi$,
and $\Omega$ and their resonances with masses up to 2.25 GeV) and 32
different meson species (including light unflavored and strange
mesons and their resonances with masses up to 2.0 GeV) as tabulated
in the PDG \cite{Amsler:2008zz}. Through baryon-antibaryon symmetry the
respective antibaryons are included. The isospin is explicitly
treated as well \cite{Bass98,Bleicher99}. For hadronic continuum excitations a string model
is used. Starting from the version 2.0, the PYTHIA is incorporated
into UrQMD in order to investigate the jet production and
fragmentation at RHIC energies \cite{Bratkovskaya:2004kv}.

\subsection{Treatments of the $\Delta$ resonance decay in UrQMD}

\begin{figure}
\includegraphics[angle=0,width=0.6\textwidth]{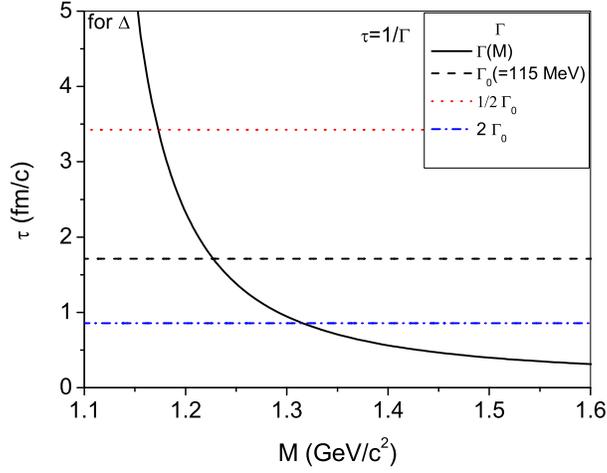}
\caption{Lifetime $\tau$ of the $\Delta(1232)$ resonance defined as
the inverse of resonance width. $1/2$, $1$, and $2$ times of
$\Delta(1232)$ width at the pole mass are selected as well as a mass
dependent one ($\Gamma(M)$). } \label{fig1}
\end{figure}

The frequently used approach to resonance lifetimes in most
transport calculations is the application of $\tau=1/\Gamma$ in
conjunction with a Monte-Carlo sampling of the exponential decay
law. The decay width $\Gamma$ of a resonance is usually either taken
to be constant ($\Gamma_0$) or mass dependent ($\Gamma(M)$)
\cite{Bass98, Bleicher99}. Fig.\ \ref{fig1} shows the lifetimes of
the $\Delta(1232)$ resonance as a function of its invariant mass.
Considering that the spectral functions might be modified (width and
the position of the mass pole) by the hot and dense nuclear medium -
e.g., recently a precision measurement of low-mass muon pairs in
158A GeV Indium-Indium collisions at the CERN-SPS was reported, and
the associated space-time averaged $\rho$-meson spectral function
shows a strong broadening, but essentially no shift in mass
\cite{Arnaldi:2006jq} - it is desirable to explore the effect of the
broadening/narrowing of resonances on pion freeze-out. Therefore,
two other mass-independent lifetimes for $\Delta(1232)$ are adopted
which correspond to the widths $1/2\Gamma_0$ and $2\Gamma_0$,
respectively (shown also in Fig.\ \ref{fig1}). The lifetimes of
other resonances are not altered in the calculations.

It should be noted that the treatment of the resonance lifetimes and
widths in a medium is a long standing problem and much argued in the
theory community. In fact, the widely used prescription for
resonance lifetimes (especially for broad resonances) in transport
models are still under debate. Especially, the recent definition of
the lifetime by the derivative of phase shift with respect to the
center-of-mass energy is thought to be more reasonable
\cite{Danielewicz:1995ay,David:1998qu,Larionov:2001zg}. However,
there still exist difficulties to adopt the ``new'' lifetime {\it
consistently} in transport models. In addition, as a result of the
scattering, the forward going part of the wave packet can suffer a
different time delay from the scattered one, which is not trivial to
implement, and, it is still not understood how to tackle the issue
of negative time delay in the real transport model calculations
\cite{Danielewicz:1995ay,Bass98}. Due to this ongoing debate, we
stick to the above mentioned schematic change by factors of 2 to see
if any effect emerges from this modification at all.

\begin{figure}
\includegraphics[angle=0,width=0.7\textwidth]{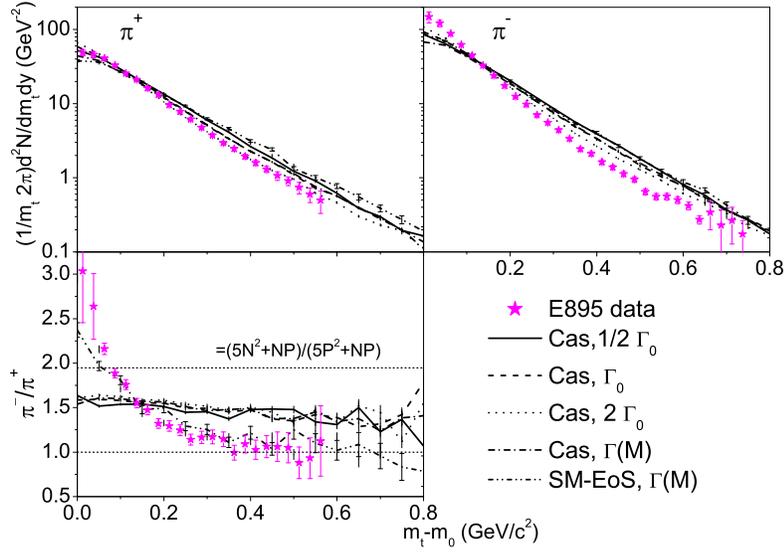}
\caption{Upper plots: Charged pion transverse mass spectra at
midrapidity ($|Y_{cm}|<0.05$) for central ($<5\%$ $\sigma_T$) Au+Au
collisions at $2$A GeV. Lower left plot: The ratio between $\pi^-$
and $\pi^+$ transverse mass spectra. The calculations with various
lifetimes in cascade mode (``Cas'') as well as in potential mode
(``SM-EoS'') are compared with the E895 data \cite{Klay:2003zf}. The
horizontal lines represent the unity and the value
$(5N^2+NP)/(5P^2+NP)$, separately (see text). } \label{fig2}
\end{figure}

Fig.\ \ref{fig2}  (upper plots) depicts transverse mass spectra of
charged pions  at midrapidity ($|Y_{cm}|<0.05$, $Y_{cm}=\frac{1}{2}{\rm
log}(\frac{E_{cm}+p_\parallel}{E_{cm}-p_\parallel})$, $E_{cm}$ and
$p_\parallel$ are the energy and longitudinal momentum of the pion
in the center-of-mass system) for central ($<5\%$
$\sigma_T$) Au+Au collisions at $2$A GeV. The lower plot shows the
ratio between $\pi^-$ and $\pi^+$ transverse mass spectra. The data
are taken from Ref.\ \cite{Klay:2003zf}. The lines correspond to
calculations with and without potentials (the ``SM-EoS'' and the
``Cas'' modes), as well as calculations with mass-dependent
lifetimes for all resonances (with respect to ``$\Gamma (M)$'') and
with three mass-independent lifetimes for $\Delta(1232)$ resonance
(with respect to $1/2\Gamma^\Delta_0$, $\Gamma^\Delta_0$, and
$2\Gamma^\Delta_0$, respectively) while other particles are not
changed. In line with the previous UrQMD calculations
\cite{Bratkovskaya:2004kv}, the $\pi$ transverse mass spectrum can
be well described by the cascade calculation, except for a slight
deviation from the data at large transverse masses. It is found that
a broader constant width of the $\Delta(1232)$ resonance makes the
spectra a little steeper since the $\Delta$ resonances decay
earlier. The mass dependence of the resonance lifetime only
influences slightly the spectra in the low transverse mass region,
which is mainly due that the $\Delta$ resonances with their
relatively small masses have long lifetimes, as a consequence, have
a higher probability to be absorbed through the $N\Delta\rightarrow
NN$ detailed balance process. However, it is interesting to see in
the lower plot of Fig.\ \ref{fig2} that the ratio between $\pi^-$
and $\pi^+$ spectra is reproduced fairly well with the ``SM-EoS''
mode, while there is no transverse mass dependence in the cascade
modes. The value $(5N^2+NP)/(5P^2+NP)$ (``$N$'' and ``$P$''
represent the initial neutron and proton numbers of HICs) of the
horizontal line is obtained if only the single pion production via
$\Delta$ resonances in nucleon-nucleon collisions is considered.
However, the effects of such as potentials, reabsorptions, and
rescatterings of pions, $\Delta$s, and other resonances will change
the value visibly. The strong transverse mass dependence of the
$\pi^-/\pi^+$ ratio observed in this plot is mainly due to the
Coulomb interaction between charged particles which has been
investigated before
\cite{Bass:1995pj,Barz:1997su,Barrette:1999ry,Li:2005zz,Li:2005gf}.

\subsection{CRAB analyzing program and the fitting process}
To calculate the two-particle correlator, the CRAB program is
adopted \cite{Pratthome}, which is based on the formula:
\begin{equation}
C({\bf k},{\bf q}) = \frac {\int d^4x_i d^4x_j g(x_i,p_i) g(x_j,
p_j) |\phi({\bf r}^\prime, {\bf q}^\prime)|^2} {{\int d^4x_i
g(x_i,p_i)} {\int d^4x_j g(x_j,p_j)}}. \label{cpq}
\end{equation}
Here $g(x_i,p_i)$ is an effective probability for emitting a
particle $i$ with 4-momentum $p_i=({E_i, \bf p}_i)$ from the
space-time point $x_i = (t_i, {\bf r}_i)$. $\phi({\bf r}^\prime,
{\bf q}^\prime)$ is the relative wave function with ${\bf r}^\prime$
being the relative position in the pair's rest frame. ${\bf q}={\bf
p}_i-{\bf p}_j$ and $\textbf{k}=(\textbf{p}_{i}+\textbf{p}_{j})/2$
are the relative momentum and the average momentum of the two
particles $i$ and $j$. Due to the quantum statistics, the correlator
is larger than unit at small $q$ for bosons, and in the absence of
strong and Coulomb final state interactions, the wave function of an
identical pair of bosons reads as
\begin{equation}
|\phi({\bf r}^\prime, {\bf q}^\prime)|^2=1+\cos(2{\bf q}^\prime
\cdot {\bf r}^\prime). \label{phifunc}
\end{equation}
And the two-particle correlator and the ``region of homogeneity''
can be directly related by a Fourier transformation. However, in
HICs, the Coulomb and strong interactions might distort the wave
function and the direct relationship between the correlator and the
region of homogeneity disappears. As a standard method, one can fit
the correlator as a three-dimensional (3D) Gaussian form under
various reference frames, in which the longitudinal comoving system
(LCMS) (or called as the ``Out-Side-Long'' system) is more
frequently adopted in recent years. We use it in the present work as
well. The corresponding 3D Gaussian correlation function  can be
expressed as

\begin{equation}
C(q_O,q_S,q_L)=K[1+\lambda {\rm
exp}(-R_L^2q_L^2-R_O^2q_O^2-R_S^2q_S^2-2R_{OL}^2q_Oq_L)].
\label{fit1}
\end{equation}
In Eq.~\ref{fit1} the $K$ is the overall normalization factor, the
$q_x$ and $R_x$ are the components of the pair relative momentum and
homogeneity length (HBT radius) in the $x$ direction, respectively.
The $\lambda$ parameter is commonly called the incoherence factor
and lies between 0 (complete coherence) and 1 (complete incoherence)
for bosons in realistic HICs. Because the parameter $\lambda$ might
be influenced by many additional factors, such as contamination,
long-lived resonances, or the details of the Coulomb modification in
the FSI, we regard it as a free parameter.
The $R_{OL}^{2}$ represents the cross-term and plays a role at large
rapidity \cite{Chapman:1994yv,Wiedemann:1999qn,lqf20062}.

However, due to the resonance decay and the space-time correlation
\cite{Heinz:1999rw,Retiere:2003kf,Lin:2002gc}, the emission function
has been found to deviate from a Gaussian form (the non-Gaussian
effect). In order to gain clearer information about the source, both
fitting the correlator to a (3-dimensional) Gaussian form with
higher orders of a harmonics and the (3-dimensional) imaging
technique \cite{Brown:1997ku,Danielewicz:2005qh} are available. The
high orders of harmonic fitting by the Edgeworth expansion was
proposed by Cs\"{o}rg\H{o} \cite{Csorgo:2000pf} and used in
experiments \cite{Adams:2004yc}, which is expressed as
\begin{equation}
C(q_{\rm inv})=K_{\rm inv}[1+\lambda_{\rm inv}\exp(-R_{\rm
inv}^2q_{\rm inv}^2)][1+\sum_{n=4,even}^{\infty}\frac{\kappa_{{\rm
inv},n}}{n!(\sqrt{2})^n}H_n(R_{\rm inv}q_{\rm inv})]. \label{Charm}
\end{equation}
Here the subscription ``inv'' means that the one-dimensional (1D)
correlation function is constructed in the invariant quantity
$q_{\rm inv}=\sqrt{(q^0)^2-(\mathbf{q})^2}$ while $R_{\rm inv}$ is
the corresponding 1D radius. $n$ is the order parameter, and the
$H_n$ reads as

\begin{equation}
H_n(z)=(-1)^n e^{z^2}\frac{d^n}{dz^n}e^{-z^2}. \label{Hnfunc}
\end{equation}

To calculate the HBT two-pion correlation, firstly, we select
central collisions at AGS energies in the UrQMD model: Au+Au at
$E_b=2$, $4$, $6$, and $8$A GeV ($<11\%$ of the total cross section
$\sigma_T$), with a rapidity cut $|Y_{cm}|<0.5$. For each case about
40 thousand events are calculated. All particles with their phase
space coordinates at their freeze-out times are put into the CRAB
analyzing program. Only the negatively charged pions are considered
during the analyzing process in this work (for each analysis, one
billion pion pairs are considered). We found that the Coulomb effect
in FSI on the HBT radii of the pion source is small
\cite{Li:2008ge}, so that we do not consider it in the calculations.

\section{HBT results}

\subsection{Gaussian fitting}

\begin{figure}
\includegraphics[angle=0,width=0.6\textwidth]{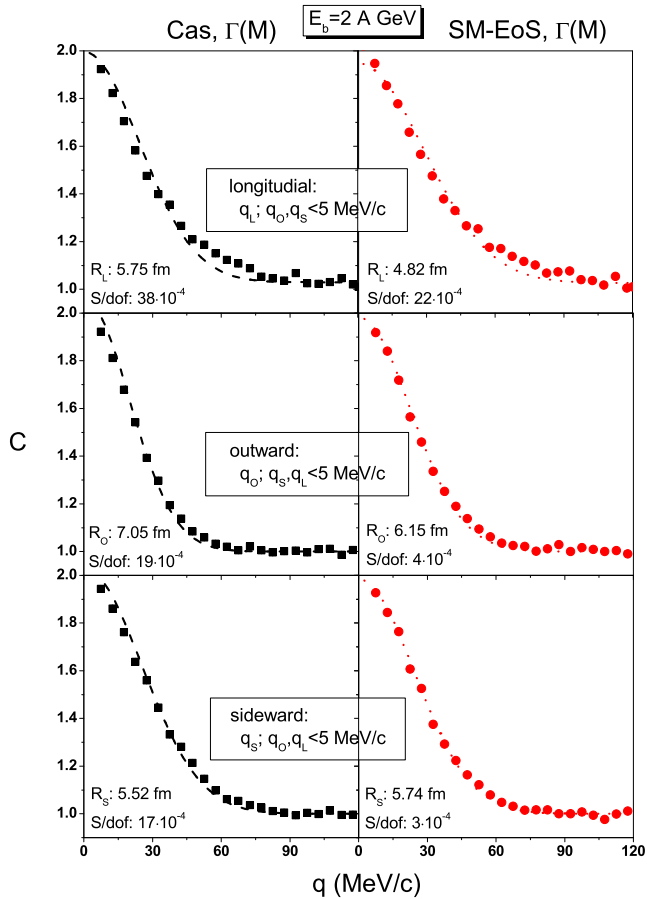}
\caption{Projections of the correlation function for negatively
charged pions in longitudinal (top plots), outward (middle) and
sideward (bottom) directions  from central Au+Au collisions at
$E_b=2$ A GeV. The results with ``Cas'' (left plots) and ``SM-EoS''
modes (right) are shown with points. The mass-dependent lifetime of
resonances is employed in the calculations. 1D Gaussian fit to each
projection is shown with lines as well.} \label{fig3}
\end{figure}

\begin{figure}
\includegraphics[angle=0,width=0.6\textwidth]{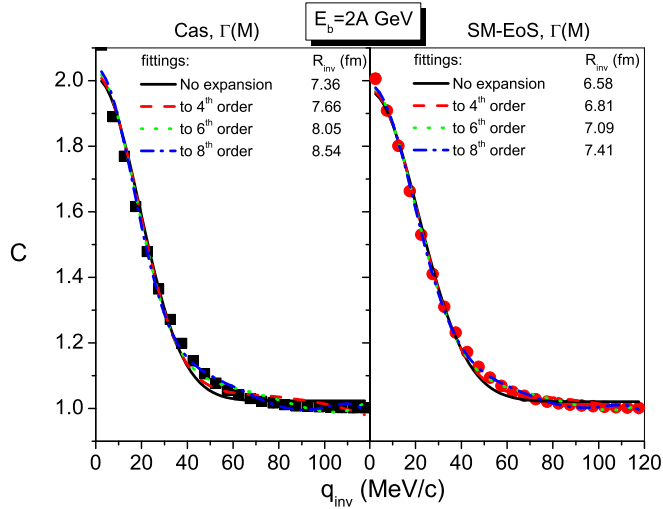}
\caption{1D correlation function for $\pi^-\pi^-$ pairs. Fittings
with the Edgeworth expansion up to the eighth order on it are also
shown. The result with cascade mode (left plot) is compared to that
with the SM-EoS mode (right). The fitting results of the radii
$R_{\rm inv}$ are listed in the plots as well.} \label{fig4}
\end{figure}

In Fig.\ \ref{fig3} we show  the projections of the calculated
correlation function of the pion source in longitudinal (top plots),
outward (middle) and sideward (bottom) directions without (``Cas'',
squares in left plots) and with (``SM-EoS'', circles in right plots)
potentials. The mass dependent lifetime of resonances is adopted.
Correspondingly, 1D Gaussian fits
($C(q_{x})=K_x[1+\lambda_{x}\exp(-R_{x}^2q_{x}^2)]$ where ``$x$''
stands for ``$L$'', ``$O$'', or ``$S$'') to them are also shown with
dashed and dotted lines, separately (the values of the HBT radius in
each projection and the corresponding reduced least square fit
$S/dof$ are also given in each plot). The fitting result shows
clearly that the non-Gaussian effect of the correlation function
exists especially in the longitudinal direction, while in other two
directions it is largely reduced. The Gaussian fitting in the
sideward direction is even better than that in the outward
direction.  We notice that in the data analysis by both the HBT
fitting and the imaging method the similar non-Gaussian effect has
been observed as well at all AGS, SPS, and RHIC beam energies
\cite{Panitkin:2001qb,Chung:2008fu,Adams:2004yc}. The non-Gaussian
character originates mainly from both the decay of long- and
intermediate-lived resonances, such as $\eta^\prime$, $\omega$,
$K^*$ mesons, and $\Delta$, $\Lambda^*$, $\Sigma^*$ baryons, and the
space-time correlation (or flow) effect
\cite{Heinz:1999rw,Retiere:2003kf,Lin:2002gc}. At AGS energies, the
contributions from meson and hyperon resonances are limited, but the
$\Delta$ resonance decay plays an important role. The origin of the
non-Gaussian effect shown in Fig.~\ref{fig3} might be clear by
comparing the HBT results with and without the mean field potential.
If comparing the left and the right plots of Fig.~\ref{fig3}
roughly, one may find that the Gaussian fitting becomes even better
with the consideration of the mean field potential.

To quantitatively analyze the non-Gaussian effect, the Edgeworth
expansions up to the 8th order on the 1D correlation function are
shown in Fig.~\ref{fig4}. It is seen that the $R_{\rm inv}$
increases with the order of the expansion on the correlation
function. In the cascade mode, the difference of the $R_{\rm inv}$
values between without expansion and with expansion up to the
$8^{th}$ order reaches $1.18$ fm, but in the SM-EoS mode the
difference is reduced to $0.83$ fm. Similarly, if one takes the same
expansions into the long-out-side projections, it is found that, 1)
in the cascade mode, the radii in longitudinal, outward, and
sideward directions change $2.2$, $0.6$, and $0.52$ fm,
respectively. 2) While in the SM-EoS mode, the changes are $1.45$,
$0.51$, and $0.38$ fm, respectively. Therefore, the non-Gaussian
effect is reduced in all directions of the correlation after
considering the mean field potential. Therefore, it is the long- and
intermediate-lived resonance decay that mainly contributes to the
visible non-Gaussian phenomenon but not the mean field.

\subsection{Transverse momentum dependence of the HBT radii}

\begin{figure}
\includegraphics[angle=0,width=0.8\textwidth]{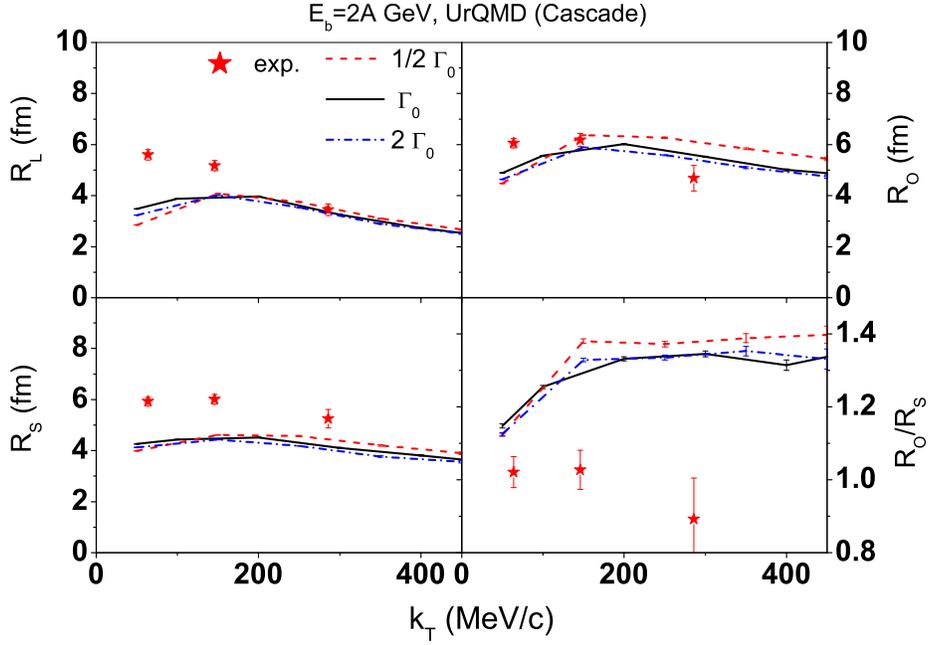}
\caption{Transverse momentum $k_T$ dependence of the HBT radii
$R_L$, $R_O$, $R_S$, and the $R_O/R_S$ ratio in central Au+Au
collisions at $E_b=2$A GeV. The cascade mode is chosen. The
mass-independent lifetime of resonances is considered, while the
width of $\Delta(1232)$ resonance is varied from $1/2\Gamma_0$,
$\Gamma_0$, to $2\Gamma_0$ (the widths of other resonances are not
changed). Experimental data are shown \cite{Lisa:2000hw} with
scattered stars. } \label{fig5}
\end{figure}

\begin{figure}
\includegraphics[angle=0,width=0.9\textwidth]{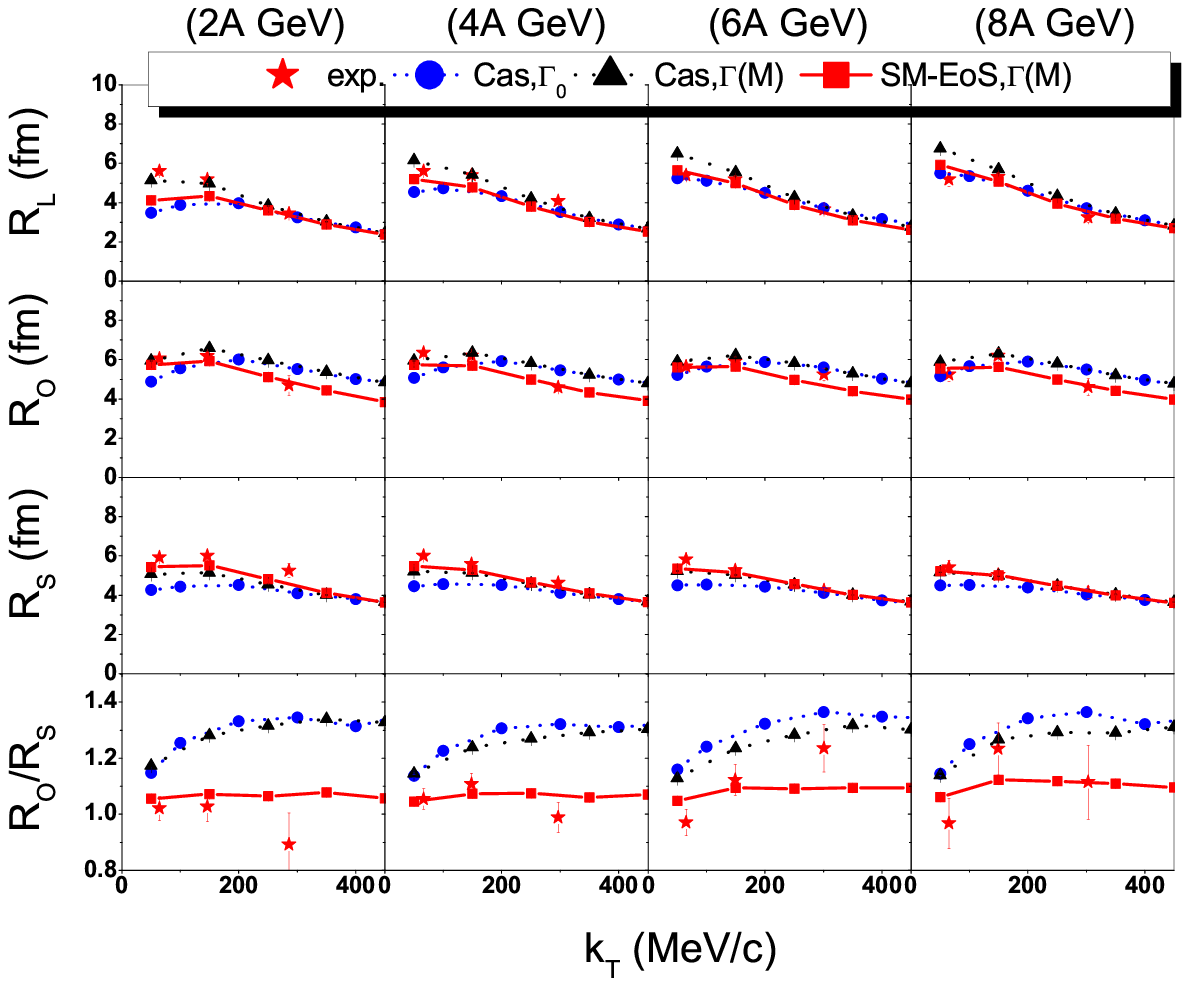}
\caption{$k_T$ dependence of HBT radii $R_L$, $R_O$, and $R_S$ as
well as the $R_O/R_S$ ratio of negatively charged pion source (from
top to bottom plots) for central Au+Au collisions at $E_b=2$, $4$,
$6$, and $8$A GeV (from left to right plots). In calculations, 1)
the``Cas,$\Gamma_0$'' represents the cascade mode with a
mass-independent lifetime of resonances. 2) ``Cas,$\Gamma(M)$''
means the cascade mode with a mass-dependent lifetime of resonances.
3) While the ``SM-EoS,$\Gamma(M)$'' represents the potential mode
with a mass-dependent lifetime of resonances. The experimental data
are taken from \cite{Lisa:2000hw}.} \label{fig6}
\end{figure}

Fig.\ \ref{fig5} gives the transverse momentum $k_T$
($\textbf{k}_T=(\textbf{p}_{1T}+\textbf{p}_{2T})/2$) dependence of
the HBT-radii $R_L$, $R_O$, $R_S$, and the $R_O/R_S$ ratio in
central Au+Au collisions at $E_b=2$A GeV. Experimental data are
taken from Ref.\ \cite{Lisa:2000hw}. The cascade calculations with
mass-independent lifetimes of resonances are shown with different
widths of the $\Delta(1232)$ resonance (the widths of other
resonances are not changed). With the default width ($\Gamma_0$), we
find that the radii $R_L$ and $R_S$ are somewhat smaller than the
data at small $k_T$. With a narrower $\Delta(1232)$ width (hence a
longer lifetime) the HBT radii at large $k_T$ are increased
slightly, and vice versa. This modification can not explain the
experimental $R_O$ and $R_S$ data, which can also be clearly seen
from their $R_O/R_S$ ratios in the right-bottom plot: the variation
of the $\Delta(1232)$ width does not help to obtain a small value of
the $R_O/R_S$ ratio as observed in the data.

Comparing the left and the right plots of Fig.~\ref{fig3} more
carefully one can find that the mean field potential enhances the
incoherence in the reaction plane (x-z plane, correspondingly, the
outward-longitudinal plane) but not in the sideward direction. Fig.\
\ref{fig6} illustrates the $k_T$ dependent HBT radii $R_L$, $R_O$,
$R_S$, and $R_O/R_S$ ratio (from top to bottom plots) of the pion
source for the energies $E_b=2$, $4$, $6$, and $8$A GeV (from left
to right plots). With the mass dependent lifetimes for all
resonances in the cascade mode (''Cas,$\Gamma(M)$''), the resonances
with their small invariant masses decay later and hence expand the
source when compared with the mass-independent lifetime of
resonances (''Cas,$\Gamma_0$''). Considering the fact that the
resonances with small invariant masses tend to produce pions with
small momenta, an increase of the HBT radii at small $k_T$ and $E_b$
is understandable. At large $k_T$ as well as at high beam energies,
this effect is reduced since less resonances with small invariant
masses contribute. It is interesting to see that the result with a
mass dependent treatment of resonance lifetimes can match the HBT
data better in the AGS energy region. However, it is also seen that
the mass dependence will increase the HBT radii with almost equal
power in outward and sideward directions and the ratio between $R_O$
and $R_S$ values remains basically unchanged. With the help of the
SM-EoS, it is interesting to see that the $R_O$ at large $k_T$ are
driven down while the $R_S$ at small $k_T$ are pulled up so that the
$k_T$ dependence of the $R_O/R_S$ ratio can be much better
described, in line with Ref.\ \cite{Li:2007yd}. This finding
supports that it is the mean field which leads to a stronger
phase-space correlation that results in a much better explanation of
the HBT time-related puzzle \cite{Li:2008ge}.

\section{Summary and outlook}
To summarize, the HBT correlation of $\pi^--\pi^-$ pairs at AGS
energies were investigated by using the UrQMD transport model and
the CRAB analyzing program. We found that a narrower (wider) width
of $\Delta(1232)$ resonance results in larger (smaller) HBT radii in
all directions at large $k_T$, which can not explain the small
experimental ratio between $R_O$ and $R_S$. Although a mass
dependent lifetime of the resonances can not resolve this problem as
well, but it pulls up the HBT ratii at small $k_T$ and hence
slightly improves the HBT radii of pions.

We observed that the $k_T$ dependent HBT radii and $R_O/R_S$ ratio
in the AGS energy region can be described fairly well with a soft
equation of state with momentum dependence. This supports the idea
that the interaction of particles in the early stage of the reaction
(leading to stronger correlation) is the key to solve the HBT
time-related puzzle.

Non-Gaussian effects are visible (although weak) in the correlation
function and might bring large uncertainties (on the order of $\pm
1$ fm) if the correlation function is fitted only by a Gaussian
form. To investigate the non-Gaussian effect, the Edgeworth
expansion was used during the fitting process. It was found that the
non-Gaussian effect is strongest in the longitudinal direction and
weakest in the sideward direction. The decay of the intermediate-
and long-lived resonances was found as the main contribution to the
non-Gaussian phenomenon, while the mean field potential did not
increase this effect.

At higher beam energies, such as SPS and RHIC, the non-Gaussian
effects as well as the broadening of vector mesons (in connection to
the NA60 results \cite{Arnaldi:2006jq}) and its influence on
$\pi^+\pi^-$ correlation deserve further investigations. In
addition, other, besides pions, (non-)identical particle
correlations deserve increased attention as they might provide
snapshots of the source geometry at different times of the reaction.
These studies are currently underway and will be addressed in a
forthcoming paper.

\section*{Acknowledgments}
We would like to acknowledge support by the Frankfurt Center for
Scientific Computing (CSC). This work was supported by the Hessian
LOEWE initiative through the Helmholtz International Center for FAIR
(HIC for FAIR).


\newpage
\begin{thebibliography}{99}

\bibitem{HBT54}R. Hanbury-Brown and R.Q. Twiss, Philos. Mag. {\bf 45}, 663 (1954); Nature (London) {\bf 178}, 1046 (1956).

\bibitem{Goldhaber60}
G.~Goldhaber, {\it et al.}, Phys. Rev. {\bf 120}, 300 (1960).

\bibitem{Bauer:1993wq}
  W.~Bauer, C.~K.~Gelbke and S.~Pratt,
  Ann.\ Rev.\ Nucl.\ Part.\ Sci.\  {\bf 42}, 77 (1992).

\bibitem{Rischke:1996em}
  D.~H.~Rischke and M.~Gyulassy,
  Nucl.\ Phys.\  A {\bf 608}, 479 (1996).

\bibitem{Soff:2000eh}
  S.~Soff, S.~A.~Bass and A.~Dumitru,
  Phys.\ Rev.\ Lett.\  {\bf 86}, 3981 (2001).

\bibitem{Lisa:2005dd}
  M.~A.~Lisa, S.~Pratt, R.~Soltz and U.~Wiedemann,
  Ann.\ Rev.\ Nucl.\ Part.\ Sci.\  {\bf 55}, 357 (2005).

\bibitem{Li:2007im}
  Q.~Li, M.~Bleicher and H.~St\"ocker,
  J.\ Phys.\ G {\bf 34}, 2037 (2007).


\bibitem{Wiedemann:1999qn}
  U.~A.~Wiedemann and U.~W.~Heinz,
  Phys.\ Rept.\  {\bf 319}, 145 (1999).

\bibitem{Bass98}S. A. Bass {\it et al.}, [UrQMD-Collaboration], Prog. Part. Nucl. Phys. {\bf 41}, 255 (1998).

\bibitem{Bleicher99}M. Bleicher {\it et al.}, [UrQMD-Collaboration], J. Phys. G: Nucl.
Part. Phys. {\bf 25}, 1859 (1999).

\bibitem{Bratkovskaya:2004kv}
  E.~L.~Bratkovskaya {\it et al.},
  Phys.\ Rev.\ C {\bf 69}, 054907 (2004).

\bibitem{Koonin:1977fh}
  S.~E.~Koonin,
  Phys.\ Lett.\ B {\bf 70}, 43 (1977).

\bibitem{Pratt:1994uf}
  S.~Pratt {\it et al.},
  Nucl.\ Phys.\ A {\bf 566}, 103C (1994).

\bibitem{Pratthome}
S. Pratt, 2000, CRAB version 3,
http://curly.pa.msu.edu/$\sim$scottepratt/freecodes/crab/home.html.


\bibitem{Li:2007yd}
  Q.~Li, M.~Bleicher and H.~St\"ocker,
  Phys.\ Lett.\  B {\bf 659}, 525 (2008).

\bibitem{Li:2008ge}
  Q.~Li, M.~Bleicher and H.~St\"ocker,
  Phys.\ Lett.\  B {\bf 663}, 395 (2008).

\bibitem{lqf20062}
  Q.~Li, M.~Bleicher, X.~Zhu and H.~St\"ocker,
  J.\ Phys.\ G {\bf 33}, 537 (2007).

\bibitem{lqf2006} Q. Li, M. Bleicher, and H. St\"ocker, Phys. Rev. C {\bf 73}, 064908 (2006).


\bibitem{Csorgo:1995bi}
  T.~Cs\"{o}rg\H{o} and B.~Lorstad,
  Phys.\ Rev.\  C {\bf 54}, 1390 (1996).

\bibitem{Bearden:2001sy}
  I.~G.~Bearden {\it et al.}  [the NA44 Collaboration],
  Phys.\ Rev.\ Lett.\  {\bf 87}, 112301 (2001).


\bibitem{Bass:1997xw}
  S.~A.~Bass {\it et al.},
  Phys.\ Rev.\ Lett.\  {\bf 81}, 4092 (1998).

\bibitem{urqmdweb1}http://th.physik.uni-frankfurt.de/$\sim$urqmd; H.~Petersen, M.~Bleicher, S.~A.~Bass and H.~St\"ocker,
  arXiv:0805.0567 [hep-ph].


\bibitem{Aichelin:1986wa}
  J.~Aichelin and H.~St\"ocker,
  Phys.\ Lett.\  B {\bf 176}, 14 (1986).

\bibitem{Aichelin:1991xy}
  J.~Aichelin,
  Phys.\ Rept.\  {\bf 202}, 233 (1991).

\bibitem{Sorge:1989dy}
  H.~Sorge, H.~St\"ocker and W.~Greiner,
  Annals Phys.\  {\bf 192}, 266 (1989).

\bibitem{Li:2005gf}
  Q.~Li, Z.~Li, S.~Soff, M.~Bleicher and H.~St\"ocker,
  J.\ Phys.\ G {\bf 32}, 151 (2006).

\bibitem{Bass:1995pj}
  S.~A.~Bass, C.~Hartnack, H.~St\"ocker and W.~Greiner,
  Phys.\ Rev.\  C {\bf 51}, 3343 (1995).

\bibitem{Li:2005kq}
  Q.~Li, Z.~Li, S.~Soff, M.~Bleicher and H.~St\"ocker,
  Phys.\ Rev.\ C {\bf 72}, 034613 (2005).



\bibitem{Li:2006wc}
  Q.~Li, Z.~Li and H.~St\"ocker,
  Phys.\ Rev.\  C {\bf 73}, 051601 (2006).

\bibitem{Li:2005zz}
  Q.~Li, Z.~Li, S.~Soff, R.~K.~Gupta, M.~Bleicher and H.~St\"ocker,
  J.\ Phys.\ G {\bf 31}, 1359 (2005).


\bibitem{Petersen:2006vm}
  H.~Petersen, Q.~Li, X.~Zhu and M.~Bleicher,
  Phys.\ Rev.\  C {\bf 74}, 064908 (2006).

\bibitem{Maruyama:1996rn}
  T.~Maruyama, K.~Niita, T.~Maruyama, S.~Chiba, Y.~Nakahara and A.~Iwamoto,
  Prog.\ Theor.\ Phys.\  {\bf 96}, 263 (1996).

\bibitem{Isse:2005nk}
  M.~Isse, A.~Ohnishi, N.~Otuka, P.~K.~Sahu and Y.~Nara,
  Phys.\ Rev.\  C {\bf 72}, 064908 (2005).

\bibitem{Hama:1990vr}
  S.~Hama, B.~C.~Clark, E.~D.~Cooper, H.~S.~Sherif and R.~L.~Mercer,
  Phys.\ Rev.\  C {\bf 41}, 2737 (1990).


\bibitem{Lisa:2000hw}
  M.~A.~Lisa {\it et al.}  [E895 Collaboration],
  Phys.\ Rev.\ Lett.\  {\bf 84}, 2798 (2000).

\bibitem{Amsler:2008zz}
  C.~Amsler {\it et al.}  [Particle Data Group],
  Phys.\ Lett.\  B {\bf 667}, 1 (2008).

\bibitem{Arnaldi:2006jq}
  R.~Arnaldi {\it et al.}  [NA60 Collaboration],
  Phys.\ Rev.\ Lett.\  {\bf 96}, 162302 (2006).

\bibitem{Danielewicz:1995ay}  P.~Danielewicz and S.~Pratt,
Phys.\ Rev.\ C {\bf 53}, 249 (1996).

\bibitem{David:1998qu}
  C.~David, C.~Hartnack and J.~Aichelin,
  Nucl.\ Phys.\ A {\bf 650}, 358 (1999).

\bibitem{Larionov:2001zg}  A.~B.~Larionov, M.~Effenberger, S.~Leupold and U.~Mosel,
Phys.\ Rev.\ C {\bf 66}, 054604 (2002).

\bibitem{Klay:2003zf}
  J.~L.~Klay {\it et al.}  [E895 Collaboration],
  Phys.\ Rev.\ C {\bf 68}, 054905 (2003).


\bibitem{Barz:1997su}
  H.~W.~Barz, J.~P.~Bondorf, J.~J.~Gaardhoje and H.~Heiselberg,
  Phys.\ Rev.\  C {\bf 57}, 2536 (1998).

\bibitem{Barrette:1999ry}
  J.~Barrette {\it et al.}  [E877 Collaboration],
  Phys.\ Rev.\  C {\bf 62}, 024901 (2000).

\bibitem{Chapman:1994yv}
  S.~Chapman, P.~Scotto and U.~W.~Heinz,
  Phys.\ Rev.\ Lett.\  {\bf 74}, 4400 (1995).

\bibitem{Lin:2002gc}
  Z.~Lin, C.~M.~Ko and S.~Pal,
  Phys.\ Rev.\ Lett.\  {\bf 89}, 152301 (2002).

\bibitem{Heinz:1999rw}
  U.~W.~Heinz and B.~V.~Jacak,
  Ann.\ Rev.\ Nucl.\ Part.\ Sci.\  {\bf 49}, 529 (1999).

\bibitem{Retiere:2003kf}
  F.~Retiere and M.~A.~Lisa,
  Phys.\ Rev.\  C {\bf 70}, 044907 (2004).

\bibitem{Brown:1997ku}
  D.~A.~Brown and P.~Danielewicz,
  Phys.\ Lett.\  B {\bf 398}, 252 (1997).

\bibitem{Danielewicz:2005qh}
  P.~Danielewicz and S.~Pratt,
  Phys.\ Lett.\  B {\bf 618}, 60 (2005).

\bibitem{Csorgo:2000pf}
  T.~Cs\"{o}rg\H{o} and S.~Hegyi,
  Phys.\ Lett.\  B {\bf 489}, 15 (2000).


\bibitem{Adams:2004yc}
  J.~Adams {\it et al.}  [STAR Collaboration],
  Phys.\ Rev.\  C {\bf 71}, 044906 (2005).


\bibitem{Chung:2008fu}
  P.~Chung and P.~Danielewicz,
  arXiv:0807.4892 [nucl-ex].


\bibitem{Panitkin:2001qb}
  S.~Y.~Panitkin {\it et al.}  [E895 Collaboration],
  Phys.\ Rev.\ Lett.\  {\bf 87}, 112304 (2001).


\bibitem{Arnaldi:2006jq}
  R.~Arnaldi {\it et al.}  [NA60 Collaboration],
  Phys.\ Rev.\ Lett.\  {\bf 96}, 162302 (2006).


\end{thebibliography}
\end{document}